\documentstyle[epsf,amssym]{report}
\renewcommand{\abstract}[1]{{\footnotesize \noindent {\bf 
Abstract}
#1 \\}}
\renewcommand{\author}[1]{\subsubsection*{#1}}
\newcommand{\address}[1]{\subsubsection*{\it#1}}
\begin{document}

\chapter*{Galactic Environments of the Sun and Cool Stars\footnote{To be published in {\it Planetary Systems -- The Long View}, eds.
L. M. Celnikier and J. Tran Than Van, Editions Frontieres, 1998}}

\author{Priscilla C. Frisch}
\address{University of Chicago, Dept. Astronomy and Astrophysics,
5640 S. Ellis Ave., Chicago, IL  60637}

\abstract{
The importance of understanding the current and historical galactic 
environments
of cool stars is discussed.  The penetration of interstellar gas into a 
stellar astrosphere
is a function of the interaction of the star with the interstellar cloud 
surrounding the
star, and this factor needs to be understood if an efficient search 
for life-bearing planets is to be made.  For the Sun, both current and 
historical 
galactic conditions are such that if a solar wind were present, it 
would have excluded  
most inflowing interstellar matter from the inner regions of the 
heliosphere for the past few million years.
Variations in heliosphere size over the recent historical path of the 
Sun are estimated, along with estimates of astrosphere sizes for 
selected nearby stars.
Considering only possible effects due to encounters with
interstellar clouds, stable planetary
climates are more likely for inner than outer planets.}

\section{Introduction}
The Sun moves through space at a velocity of about 17 pc per million years.  This motion, combined with 
interstellar cloud 
motions driven by stellar evolution, yield a constantly changing 
galactic environment 
for the Sun and solar system.  This environment affects the 
interplanetary environments of both outer and inner planets in the 
solar system, including Sun--Earth coupling mechanisms.
By analogy, the interactions between other cool stars and the galactic 
environment of that star needs to be understood as part of the 
process of identifying planets conducive to ``higher'' life forms.  

\begin{figure}[ht]
\vspace*{4in}
\includegraphics{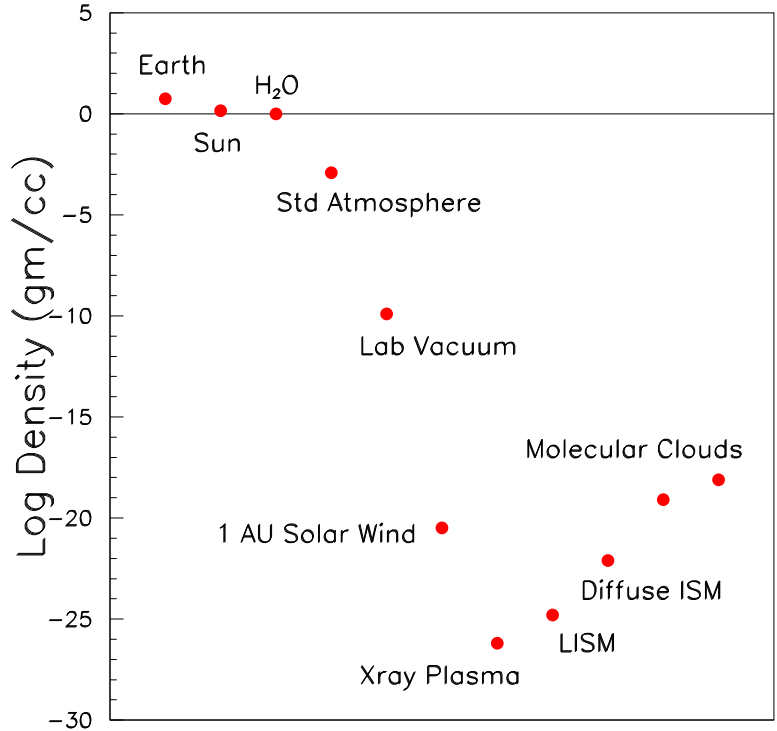}
\caption{Typical densities for material in our Galaxy.
\label{densities}}
\end{figure}

The interstellar cloud surrounding the Sun at this time (known as the 
``local interstellar cloud'', LIC), is warm, low density, and partially 
ionized: T$\approx$7,000 K, n(H$^{\rm \circ})$$\approx$0.2 cm$^{-
3}$, and n(e$^{-}$)$\approx$0.1 cm$^{-3}$.  The standard assumption 
for diffuse interstellar clouds is that n(p$^{+}$)=n(e$^{-}$).  On the 
scale of 
typical
cloud densities, the LIC is rather tenuous, and notably lower density 
than
the 1 au solar wind density (see Fig. \ref{densities}).  This accounts for
the ability of the solar wind today to exclude most interstellar 
material from 1 au.

In this paper the basis for understanding the relation between the
properties of stellar wind envelopes around cool star systems and 
the physical
properties of the surrounding interstellar clouds are examined.  The 
author believes that the historical galactic environment of a star 
would have a direct impact on the stability of
planetary atmospheres, and therefore on the distribution of 
intelligent life forms.  This conclusion rests partly on the 
observation that the solar system has been in a region of space
virtually devoid of interstellar matter over the past several million 
years \cite{fy86}, \cite{journey}.

Additional reviews on interstellar matter (ISM) within the solar
system can be found in the book {\it The Heliosphere in the Local
Interstellar Medium} \cite{rudi}.  For more information on the
properties of
local ISM (LISM) see \cite{fr95}.  For more information 
on nearby G-star
space motions and environments, and the use of astrospheres as a 
test for interstellar pressure, see \cite{fr93}.

\section{Penetration of ISM into Heliosphere\label{pui}}

Heliosphere studies offer an opportunity to study the impact of 
interstellar material on the interplanetary medium.  Based on 
spacecraft data, we know 
that neutral interstellar hydrogen
flowing through the heliopause region is the source of
several particle populations observed in both the inner and outer 
heliosphere, including energetic ions captured in the terrestrial
magnetosphere.  Also, interstellar dust is observed within the solar 
system.  Since the heliosphere is the only system where we are 
currently able to
evaluate the interplanetary environment in terms of the properties 
of the surrounding interstellar cloud, data on these populations are 
now briefly summarized.  Heliosphere dimensions change as a phase 
of the solar cycle; therefore both the 
interaction 
of interstellar hydrogen with the solar system, and the distribution 
of the
daughter pickup ion and anomalous cosmic ray populations, are 
solar-cycle dependent.  Hence for external cool stars, where stellar 
winds may not mimic the solar wind, the distribution of ISM and 
interaction products within the astrosphere may
differ from those in the heliosphere.

Neutral interstellar hydrogen interacts weakly in the heliopause 
region, with, for example, $\sim$20\% of the incident particles lost through charge 
exchange with interstellar protons in the heliosphere nose region,
\cite{fahr}, \cite{izm}.  However, most
of the incident interstellar H$^{\rm \circ}$ penetrates to the inner 
solar system, 2--5 au from the Sun.  Hydrogen ionization rates 
evaluated at 1 au yield 85\% ionization by charge-exchange with 
solar wind protons, and 15\% ionization by photoionization, 
depending somewhat on the phase of the
solar cycle \cite{rucinski}.  Helium destruction is dominated by 
photoionization.  
Observations of the resonance fluorescence of solar radiation is 
observed in the Ly$\alpha$ radiation of H$^{\rm \circ}$ and the 584 
A line of He$^{\rm \circ}$, \cite{quemerais}, 
\cite{lalhalpha}, \cite{ajello}, \cite{adamsfrisch} and references 
therein.  The 
main fluorescence for H$^{\rm \circ}$ originates from a region 3--5 
au from the Sun, while most of the He$^{\rm \circ}$ fluorescence 
arises from the inner $<$0.5 au.

Once ionized, interstellar hydrogen is bound to the 
outward 
moving solar wind by the Lorentz force, becoming
pickup ions.  Heavier elements such as He, C, O, N, Ne are
also observed, \cite{gloeckler} and references therein.  As a 
measure of the importance of pickup ions on solar
wind dynamics, the partial pressures of the pickup protons and solar 
wind protons are equal at about 6 au, with the pickup proton partial
pressure surpassing the solar wind partial pressure by orders of 
magnitude external to that location \cite{whang}.  

In the region of the termination shock of the solar wind (where the
solar wind goes from being supersonic to subsonic), the pickup ions 
are accelerated to MeV energies, and they become the anomalous 
cosmic 
ray component \cite{cummings}.  

Heavy N, O, C, Ar, Ne, He ions from the anomalous cosmic ray 
populations 
become trapped in the terrestrial magnetosphere, leaking out over
a period of about two weeks \cite{mewaldt}, \cite{tylka}.

Interstellar dust grains with a mean mass of 3 10$^{-13}$ g, 
corresponding to a mean radius of 0.31 $\mu$m for density 2.5 gr 
cm$^{-3}$ have been observed by the Ulysses and Galileo satellites 
within the orbit of Jupiter
\cite{baguhl}.  Smaller charged grains are excluded at the 
heliopause and by the solar wind, but the larger grains penetrate 
more freely.

\section{Solar and Cloud Motions\label{apex}}

The discussions in this paper are based on solar and stellar motions, 
referred to the Local
Standard of Rest (LSR) velocity frame.\footnote{The LSR is the 
velocity frame of reference
in which the velocities of nearby stars average to zero.}  The Sun is 
assumed to have
a velocity of 16.5 km s$^{-1}$ through the LSR, directed towards 
galactic coordinates
l=53$^{\rm \circ}$ and b=+25$^{\rm \circ}$.  Because the orbit of the 
Sun is
inclined with respect to the galactic plane, the Sun oscillates through 
the plane with
a period of $\sim$66 million years, an amplitude of $\sim$80 pc, and 
the last galactic plane crossing $\sim$21 Mys ago
\cite{bash}.  The amplitude of oscillation about the galactic plane is 
not large enough to carry the Sun out of the H$^{\rm \circ}$ disk of 
the plane.

\begin{figure}[ht!]
\label{fig_ism}
\vspace*{4in}
\includegraphics{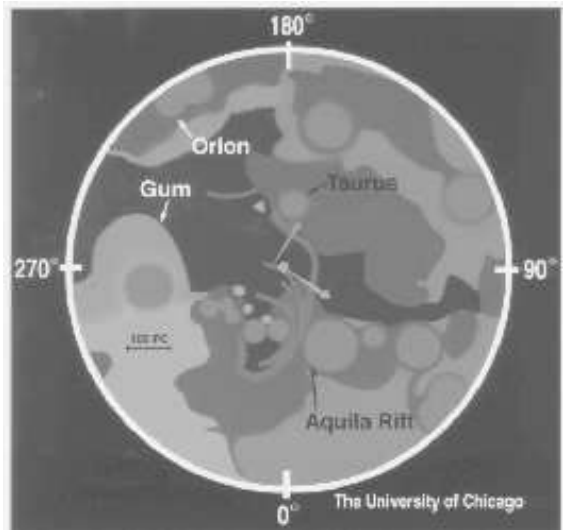}
\caption{Plot of molecular and diffuse interstellar clouds
within 500 pc of the Sun.  The view point is from the North
Galactic Pole looking down on the galactic plane, and the coordinates
are galactic coordinates.  The LSR motions of
the Sun and interstellar cloud surrounding the Sun are both shown
as arrows.  For more information, see Frisch 1998.}
\end{figure}

Five million years ago the Sun was
located at a distance of $\approx$83 pc from the current position 
of the Sun, and $\approx$35 pc below the galactic plane.  The 
Sun is emerging from a region of space that is deficient in interstellar 
matter, \cite{fy86}, \cite{journey}.  This is
seen in Fig. \ref{fig_ism}.  X-ray data indicate that the nearest 
portions of this region are deficient in H$^{\rm \circ}$, and
evidently filled with hot plasma with T$\sim$10$^{6}$ K, n(e$^{-}$)$\sim$0.005 cm$^{-3}$\footnote{This region is also known 
as the
interior of the local bubble.} \cite{cox}.  For millions of years, this hot 
plasma has constituted the galactic environment of the Sun.

The bulk motions of the interstellar clouds surrounding the
Sun are approximately perpendicular to the solar motion in the LSR, 
so that in effect
this cloud system is sweeping down on the solar system, \cite{issi}.
Based on HST observations towards Sirius, 
there is a second
interstellar cloud, in addition to the LIC cloud, towards this nearby 
star, at d=2.7 pc \cite{sirius}.  
This second cloud (which is blue-shifted relative to the LIC, hence the
term ``blue-shifted'') has n(e$^{-}$)=0.46 cm$^{-3}$, T=3600 K, and 
a relative velocity with respect to the Sun of 10 km s$^{-1}$ 
\cite{gry}.  This blue-shifted cloud is a somewhat denser diffuse 
cloud than the LIC.  
Since this cloud is seen in front of two stars at the positions 
(l,b)=(227$^{\rm \circ}$, --9$^{\rm \circ}$) and (240$^{\rm \circ}$, 
--11$^{\rm \circ}$), compared to the anti-apex direction of 
(233$^{\rm \circ}$, --25$^{\rm \circ}$), the Sun is likely to have 
encountered this
cloud sometime within the past $\sim$250,000 years.  This cloud 
provides the second 
historical environment of the Sun that will be considered.

\section{Heliospheres and Astrospheres\label{hsas}}

The fundamental variable governing the penetration of interstellar
matter into the region of space occupied by a stellar wind is the size 
of the 
heliosphere/astrosphere\footnote{The heliosphere is the domain of
space occupied by the solar wind.  It is bounded by the heliopause, 
a 
stagnation surface between solar wind and interstellar plasmas.  By 
analogy, the astrosphere is the similar domain for external stars.}.
In this section, a simple effort is made to predict the response 
of the
heliosphere to different types of interstellar clouds.  This discussion 
also applies to the sizes of astrospheres formed around cool stars 
with stellar
winds of similar properties to those of the solar wind. 
These predictions are made based on the discussions of
Holzer (1989) for the heliopause radius as a 
function of the 
equilibrium
between the ram pressure of the solar and interstellar winds.  This 
equation yields a simplified approximation to the detailed 
predictions of two and three-dimensional kinetic, multi-fluid, and 
magneto-hydrodynamic models of the heliosphere
\cite{baranov}, \cite{zank}, \cite{linde}.  The pickup ion and 
anomalous cosmic ray pressures are not included in this equation. 
However this approximation does give us a 
basis for understanding how
different cloud types might affect the properties of our heliosphere 
relative to each other.

\begin{equation}
\label{eq1}
\frac{P_{SW}}{R^{2}_{A}} = \frac{\alpha B^{2}}{8 \pi} + 2 \beta n_{e} 
k T + \gamma n_{e} m_{H} u^{2} + \delta \beta n_{H} k T + \delta 
\gamma n_{H} m_{H} u^{2} + 
\epsilon P_{CR}
\end{equation}

The variables  $P_{SW}$ and  $R_{A}$ are the solar wind 
ram pressure at 1 au and the heliopause radius (in units 
au), 
respectively.  The solar wind parameters used here are a density of 7.63 
particles cm$^{-3}$ and a velocity of 440 km s$^{-1}$, based on 14-year 
average values determined from the Voyager spacecraft 
\cite{whang}, giving a ram pressure of $P_{SW}$=2.5 10$^{-8}$ dynes 
cm$^{-2}$.  $B^{2}$ is the strength of the
interstellar magnetic field; $n_{e}$ is the interstellar electron density; 
$n_{H}$ is the 
density of interstellar H$^{o}$; T is the temperature (K); m$_{H}$ is 
the mass of the
hydrogen atom; k is the Boltzman constant; u is the relative velocity of 
the Sun
and the surrounding interstellar cloud; and $P_{CR}$ is the pressure 
of cosmic rays.
The factors of $\alpha$, $\beta$, $\gamma$, $\delta$, and $\epsilon$ 
are, respectively, assumed to be 2.5, 1,1, 0.2, 0.23 (Holzer 1989). 
The constant $\alpha$ represents the amplification of the interstellar magnetic 
field by interaction with the heliosphere; the constants $\beta$ and 
$\gamma$ relate to the flow of interstellar matter around the 
heliosphere; $\delta$ and $\epsilon$ are the fractions of interstellar 
H$^{\rm \circ}$ and galactic cosmic rays
excluded from the heliosphere, respectively.
The distance of the
heliopause from the Sun will vary according to the properties of the 
cloud surrounding
the solar system.  

The distribution of interstellar matter within the heliosphere depends on the dimensions 
of the heliosphere; therefore the expected heliopause distance is 
calculated for encounters with a 
sample of different types of interstellar clouds (Table 
\ref{HP}).  The main value of these estimates is comparative, in that 
they
illustrate the response of the heliosphere to encounters with clouds 
of different properties.  True two-dimensional and three-dimensional 
heliosphere models now exist that provide more accurate estimates 
of heliosphere dimensions (see the references above).  The salient 
properties of these results  are that 
all types of interstellar cloud types considered, except for the diffuse 
LIC, `blue-shifted' cloud, and 
hot plasma, give heliopause dimensions significantly smaller than 
found today.  This result derives partly from the assumption that
most of the clouds are at rest in the LSR, so that the relative Sun-
cloud
velocity is 17 km s$^{-1}$.  The high-velocity shock front, likewise, 
yields a smaller heliosphere, 
but the effect would be transient since such a shock front
would pass over the heliosphere in less than one year.  

The two historical solar environments identified in Section \ref{apex} 
are a hot plasma and a slightly denser diffuse cloud.  In the hot 
plasma case (column 4, Table \ref{HP}) the heliosphere would be 
slightly larger (25\%) than today.  A diffuse cloud blue-shifted with 
respect to the LIC has been observed by the Hubble Space Telescope 
towards Sirius and $\epsilon$ CMa, \cite{sirius}, \cite{gry}.  
Assuming that n(e$^{-}$)$\approx$n(H$^{\rm \circ}$),
the heliosphere would have been approximately the same size as 
today, since the larger densities are offset by the lower temperature 
and relative velocity (column 2, Table\ref{HP}).  The other clouds 
considered in the table were assumed to be at rest in the LSR, so that 
the larger densities effectively compressed the heliosphere to much 
smaller dimensions than today.  

The solar wind flux used in Table \ref{HP} is based on a 14 year 
average 
value.  During periods such as the Maunder minimum,
when sunspots virtually disappeared, 
solar wind properties were different, \cite{maunder},
indicating that the size of external cool star astrospheres will
be expected to fluctuate with starspot activity.

\begin{table}[h]
\label{HP}
\begin{center}
{
\caption{Heliopause Radius versus Cloud Type}
\begin{tabular}{lccccc ccccc}
\\
\hline
\\
Variable&Diffuse&Diffuse&Shock&Hot&10$^{2}$&10$^{3}$&10$^{4}$& 10$^{5}$&H II\\
&LIC&Blue&Front&Plasma&&&&&&\\
Cloud No.&1&2&3&4&5&6&7&8&9&\\
\\
\hline
\\

u (km s$^{-1}$)&26&10&100&17&17&17&17&17&17\\
n$_{H}$ (cm$^{-3}$)&0.2&0.46&0.8&0.0003&10$^{2}$&10$^{3}$&10$^{4}$&10$^{5}$&0 \\
n$_{e}$ (cm$^{-3}$)&0.1&0.46&0.4&0.005&0.01&0.1&1&10&10\\
T$^{\rm \circ}$&7,000&3,600&10$^{5}$&10$^{6}$&100&50&20&20&10,000\\
B ($\mu$G)&2&2&6&0&5&5&5&5&0\\
R$_{A} $ (au)&106&118&15&134&16&5&2&0.5&18\\

\\
\hline
\hline
\end{tabular}
}
\end{center}
\end{table}
\normalsize
\noindent

\section{Historical Environments of Nearby Cool Stars Traversing 
Low Density Regions for the Past 5 Myrs}

A sensible search for external planets with stable climates must be 
based on an understanding of the galactic environment of that star.  
Past studies of the historical environments of nearby stars include
a search for star-cloud encounters as a basis for understanding
the formation and equilibrium of dusty circumstellar disks around 
A-stars, \cite{whitmire}, \cite{lissauer}, and a study of single 
nearby cool stars with historical
space trajectories through regions of space
with low average spatial densities so that the star is unlikely to have 
encountered significant interstellar clouds over the past 5 Myrs 
\cite{fr93}.  Main sequence stars spend typically 3\% of their 
lifetimes inside of interstellar clouds \cite{lissauer}.  A list of a 
subset of G-stars with historical galactic environments
in low density regions of space
is given in Table \ref{AP}, along with an estimate of the astropause 
radius for each star based on the space velocity of the star given
in \cite{fr93}.  Column 1 gives the HD number of the star; columns 2--4
give the galactic coordinates and distance; V gives the velocity
of the star through the LSR. The Cloud No. column refers to the
cloud numbers in Table \ref{HP}, and the last column gives the
astropause radius predicted assuming the cloud is at rest in the LSR
with a cloud type from column 6.  The listed stars were selected as stars which
are not likely to have encountered interstellar clouds over the past
several million years.  The cloud type assumed for the 
surrounding 
interstellar cloud is also listed in the table, where the cloud type is
based on an informed guess as to the properties of the interstellar 
cloud surrounding each star.  The number listed in column 6 gives
the number of the cloud type from Table \ref{HP}.

\begin{table}[h]
\label{AP}
\begin{center}
{
\caption{Astropause Radii for Selected G-Stars}
\begin{tabular}{lcccccc} 
\hline
\\
Star&l&b&r&V&Cloud&R$_{A}$\\
HD&Deg.&Deg.&pc&km s$^{-1}$&No.&AU\\
\hline
\\
1461&101&--69&19&23&1&66\\
12235&155&--55&27&33&1&48\\
13421&155&--49&30&39&1&41\\
14412&214&--70&12&34&1&47\\
14802&209&--69&12&7&1&137\\
38529&205&--13&34&22&4&69\\
48938&237&--13&17&35&4&46\\
50692&191&13&19&25&4&62\\
84737&173&50&13&26&4&60\\
96700&278&28&20&29&1&54\\
102438&288&30&16&11&1&112\\
126053&348&55&17&38&1&42\\
147513&342&7&15&23&2&66\\
175225&83&21&24&25&1&62\\
186760&91&16&21&15&1&92\\
199960&45&--31&28&12&2&107\\
217014&91&--35&18&27&1&58\\
\hline
\\
\end{tabular}
}
\end{center}
\end{table}

\section{Closing Comments}

Based on observations of interstellar matter within our solar system, 
we can say with confidence that the interplanetary environments of 
external planetary systems are a function of the interaction of each 
star with the interstellar cloud surrounding that star.  Factors that 
regulate this interaction will be the stellar wind flow, and the  cloud 
density, temperature, and velocity relative to the star.  In turn, the 
interplanetary environment will affect the planetary atmospheres 
and
climate.  A discussion of climate variations due to encounters with
interstellar clouds is outside of the scope of 
this review.  But it is worth noting that the climates of outer planets will be 
more sensitive to perturbations by interstellar gas than will be
the climates of inner planets.  For instance, in the solar system
today the densities of the solar wind and incident interstellar gas
are equal at the orbit of Jupiter.  Therefore, considering
only the interstellar cloud encounter aspect, stable planetary climates
are more likely for inner than outer planets.
These factors need to be considered if a sensible search
for extraterrestrial life is to be conducted.

\end{document}